# Daily growth rate of scientific production on Covid-19. Analysis in databases and open access repositories

@torressalinas -Universidad de Granada

## ENGLISH ABSTRACT AND MAIN FIGURES

The scientific community is facing one of its greatest challenges in solving a global health problem: COVID-19 pandemic. This situation has generated an unprecedented volume of publications. What is the volume, in terms of publications, of research on COVID-19? The general objective of this research work is to obtain a global vision of the daily growth of scientific production on COVID-19 in different databases (Dimensions, Web of Science Core Collection, Scopus-Elsevier, Pubmed and eight repositories). In relation to the results obtained, Dimensions indexes a total of 9435 publications (69% with peer review and 2677 preprints) well above Scopus (1568) and WoS (718). This is a classic biliometric phenomenon of exponential growth (R2 = 0.92). The global growth rate is 500 publications and the production doubles every 15 days. In the case of Pubmed the weekly growth is around 1000 publications. Of the eight repositories analysed, Pubmed Central, Medrxiv and SSRN are the leaders. Despite their enormous contribution, the journals continue to be the core of scientific communication. Finally, it has been established that three out of every four publications on the COVID-19 are available in open access. The information explosion demands a serious and coordinated response from information professionals, which places us at the centre of the information pandemic.

**Figure 1.** Number of accumulated publications on COVID-19 in *Dimensions* classified by source (repositories and journals)

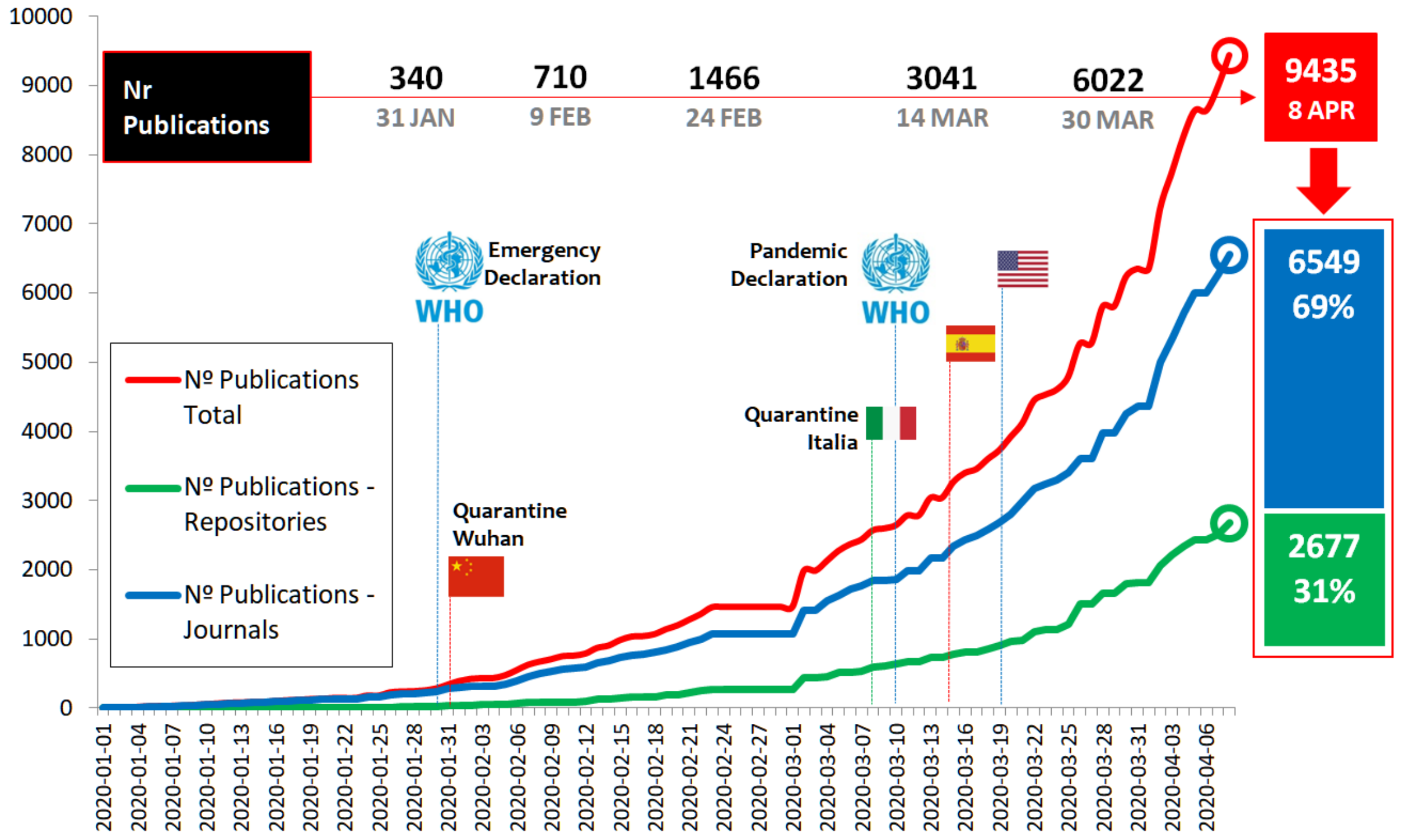

**Table 1.** Publications on COVID-19 in the Pubmed database through the records collected in Dimensions and LitCovid

| | Nr Publications | | | Weekly growth | |
|---|---|---|---|---|---|
| **Semana** | **Nr Pubs. *Pubmed*** Source= LitCovid | **Nº Pubs. *Pubmed*** Source= Dimensions | **Diferencia a favor de Dimensions** | **Weekly growth *Pubmed*** Source= LitCovid | **Weekly growth *Pubmed*** Source= Dimensions |
| 20 January - 26 Enero | 13 | 68 | 81% | | |
| 27 January - 02 February | 45 | 103 | 56% | + 246% | + 51% |
| 03 February - 09 February | 102 | 172 | 41% | + 127% | + 67% |
| 10 February - 16 February | 116 | 194 | 40% | + 14% | + 13% |
| 17 February - 23 February | 139 | 235 | 41% | + 20% | + 21% |
| 24 February - 01 March | 173 | 300 | 42% | + 24% | + 28% |
| 02 March - 08 March | 263 | 284 | 7% | + 52% | - 5% |
| 09 March - 15 March | 266 | 358 | 26% | + 1% | + 26% |
| 16 March - 22 March | 406 | 562 | 28% | + 53% | + 57% |
| 23 March - 29 March | 499 | 539 | 7% | + 23% | - 4% |
| 30 March - 5 April | 985 | 1029 | 4% | + 97% | + 91% |
| 6 April - 12 April | 915 | *317 | -- | + 1% | -- |
| **Totals** | **3922** | **3844** | -- | -- | -- |

*Data from 6-8 April

**Figure 2.** Number of accumulated publications on COVID-19 in eight different repositories

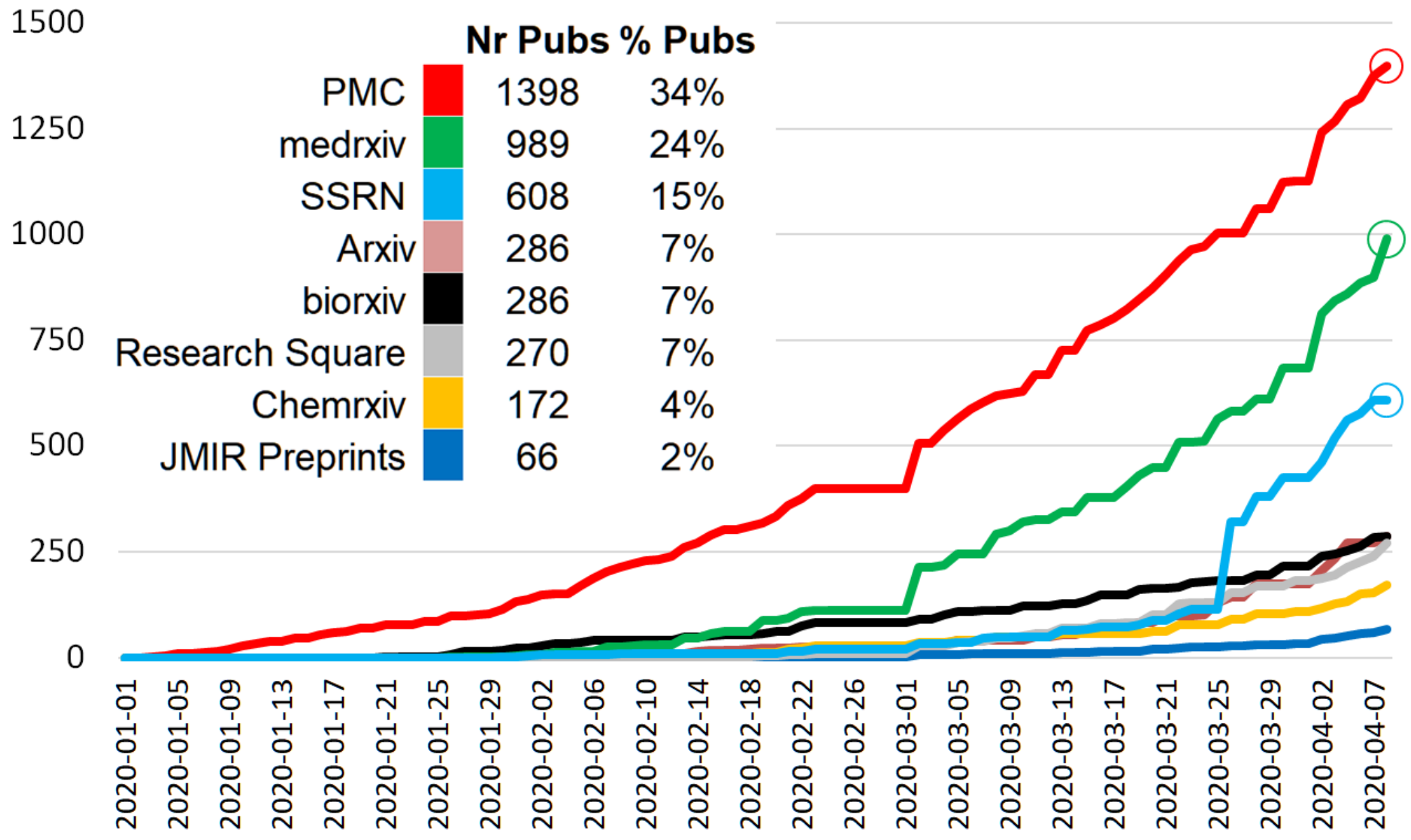

# *Ritmo de crecimiento diario de la producción científica sobre Covid-19 Análisis en bases de datos y repositorios en acceso abierto*

*Daily growth rate of scientific production on Covid-19. Analysis in databases and open access repositories*

**Daniel Torres Salinas**
Universidad de Granada, Departamento de Información y Comunicación, Medialab UGR, Unidad de Excelencia Iberlab y Ec3metrics spin off. Campus Cartuja s/n
Email- torressalinas@go.ugr.es – ORCID: https://orcid.org/0000-0001-8790-3314

**Resumen**: La comunidad científica se encuentra ante uno de sus mayores retos para resolver un problema sanitario de alcance global como es la pandemia del COVID-19. Esta situación ha generado un volumen de publicaciones sin precedentes pero ¿cuál es el volumen, en términos de publicaciones, de la investigación sobre el COVID-19?. Conseguir una visión global sobre el crecimiento diario de la producción científica sobre el COVID-19 en diferentes fuentes de información (Dimensions, Web of Science Core Collection, Scopus-Elsevier, Pubmed y ocho repositorios) es el objetivo general de este trabajo de investigación. En relación a los resultados obtenidos Dimensions indexa un total de 9435 publicaciones (69% con revisión por pares y 2677 preprints) muy por encima de Scopus (1568) y WoS (718). Nos encontramos ante un fenómeno clásico biliométrico de crecimiento exponencial (R2 =0,92). El ritmo de crecimiento a nivel global es de 500 publicaciones diarias en los últimos días y la producción de duplica cada 15 días. En el caso de Pubmed el crecimiento semanal se sitúa en torno a las 1000 publicaciones, tanto en Dimensions como en Litcovid. De los ocho repositorios analizados se sitúa a la cabeza Pubmed Central, Medrxiv y SSRN, a pesar de la enorme contribución de los mismos las revistas siguen siendo el núcleo de la comunicación científica. Finalmente se ha establecido que tres de cada cuatro publicaciones sobre el COVID-19 están disponibles en acceso abierto. Estas cifras exigen una respuesta de los profesionales de la información ante una explosión de información sin precedentes que nos sitúa en el centro de la pandemia informativa.

# *Ritmo de crecimiento diario de la producción científica sobre Covid-19 Análisis en bases de datos y repositorios en acceso abierto*

## 1. Introducción

La comunidad científica se encuentra ante uno de sus mayores retos para resolver un problema sanitario de alcance global como es la pandemia del COVID-19. La situación en la que nos encontramos inmersos requiere un esfuerzo científico colectivo que se refleja diariamente en la publicación de cientos de documentos y recursos de todo tipo (artículos, preprints, guías clínicas, protocolos, etc.) que involucran a todas las áreas del conocimiento. Probablemente estamos asistiendo a la mayor concentración de recursos científicos para la resolución de un problema concreto superando con creces otros precedentes como pudieran ser el proyecto Manhattan o la misión Apolo. En este contexto la eficacia del sistema de comunicación y publicación científica y sus diferentes elementos (revistas, revisores, bases de datos, repositorios) está siendo puesta a prueba ante una cantidad de conocimiento generado en un breve lapso de tiempo que no tiene precedentes (**Kupferschmidt**, 2020).

El primer problema inmediato al que ha debido enfrentarse el universo de la publicación es la avalancha de artículos y prepints y la necesitad que éstos sean accesibles. Una de las respuestas colectivas por parte de las editoriales ha sido la creación de centros de recursos que unifican en una única web y en acceso abierto todo aquello que se va publicando sobre el COVID-19. Así, las multinacionales de la edición de cómo *Elsevier*, *Springer* o *Emerald* han adoptado esta política. Las grandes revistas científicas, especialmente las de biomedicina, también están haciendo frente a este escenario compartiendo todas sus publicaciones. Así lo están haciendo JAMA, BMJ, SCIENCE, Oxford, Cambridge o New England. Asimismo los repositorios, por su rapidez y eficacia en transmitir la información científica, están en el centro de las miradas y la mayor parte facilitan en sus páginas de entrada consultas rápidas a sus preprints, como ocurre en Arxiv, y otros, como Zenodo, han creado comunidades que recopilan los trabajos más relevantes. Ante esta avalancha surge una pregunta básica: ¿cuál es el volumen, en términos de publicaciones, de la investigación sobre el COVID-19?

Debido a la multidisciplinaried del frente de investigación del COVID-19 y la multitud de fuentes de datos una cuantificación urgente y una caracterización global, aunque sea de carácter descriptivo, ayudaría a los profesionales de la información a comprender y visualizar un fenómeno informativo al que tenemos que enfrentarnos en los próximos meses. Si bien es cierto que ya se están realizando los primeros análisis bibliométricos éstos están centrados en analizar las fuentes tradicionales como *Web of Science*, *Scopus* o *Pubmed* y en describir a los productores de las publicaciones (por ejemplo **Chahrour**, 2020; **Hossain**, 2020; **Alba**, 2020). Sin embargo estos trabajos obvian la importancia de otras fuentes como son los repositorios en sus cuantificaciones y ofrecen una visión fragmentada del fenómeno. Conseguir una visión global sobre el crecimiento diario de la producción científica sobre el COVID-19 en diferentes fuentes es el objetivo general de este trabajo, más específicamente se han establecido dos objetivos:

1) Cuantificar de forma global cuál es el volumen de la producción científica sobre el COVID-19 en diferentes bases de datos multidisplinares y especializadas como *Web of Science*, *Scopus*, *Dimensions* y *Medline* y determinar su crecimiento diario.
2) Cuantificar del mismo modo el número de preprints diarios que se publican en los distintos repositorios y describir el fenómeno del acceso Abierto

Los resultados que se alcacen en este trabajo no solo tienen un valor descriptivo de un fenómeno informativo singular, si no que nos permitirá a los profesionales de la información tener un mapa objetivo y global de aquellas de las fuentes de información y bases de datos más útiles para enfrentarnos al COVID-19 y asesorar a nuestros investigadores.

## 2. Material y métodos

Para la realización de este trabajo se han utilizado diferentes recursos. El primero de ellos es *Dimensions*, una base de datos multidisciplinar que integra diferentes recursos (publicaciones, datos, ensayos clínicos) en un único punto. Indexa contenido de revistas científicas, bases de datos como *Pubmed* y preprints procedentes de ocho repositorios en acceso abierto. No se ha consultado *Dimensions* directamente si no que se ha hecho del dataset público con las referencias a todas las publicaciones sobre el COVID-19 (https://tinyurl.com/qqzncmv). Dicho fichero fue sometido a un proceso de normalización en diversos campos. También se ha consultado las bases de datos *Web of Science Core Collection* completa y la base de datos *Scopus* de *Elsevier*. La recuperación de información en *Dimensions*, WoS y *Scopus* se ha realizado mediante la ecuación de búsqueda ofrecida por *Dimensions* en su dataset, es la siguiente:

*PUBLICATION YEAR: 2020; FULL DATA SEARCH: "2019-nCoV" OR "COVID-19" OR “SARS-CoV-2” OR (("coronavirus" OR "corona virus") AND (Wuhan OR China))*

Tal y como puede observarse las publicaciones se refieren exclusivamente al COVID-19 y no se incluyen otros términos tangenciales. Se ha hecho, por tanto uso de la misma ecuación en las tres bases de datos. Las consultas se realizaron durante los días 11 y 12 de Abril. Una cuarta fuente que se ha empleado es *LitCovid* una web centralizada de publicaciones sobre el COVID-19, curada y mantenida por el NIH (**Chen, Allot & Lu**, 2020). Es una fuente sobre publicaciones verificadas del COVID-19 sin los errores que pueda provocar una consulta mal definida. Esta fuente nos ayudará a cuantificar de forma realista las publicaciones indexadas en *Pubmed* y ayudará a verificar las publicaciones de *Pubmed* indexadas en *Dimensions*. Los resultados e indicadores que se presentan son indicadores básicos de producción, se han considerando en todos las indicadores de cualquier tipología documental.

## 3. Resultados

### 3.1. Bases de datos científicas

A la hora de resolver un problema bibliométrico el primer lugar donde acudir son las grandes bases de datos científicas WoS y *Scopus*. En la Tabla 1 se presenta los resultados de la búsqueda. *Web of Science* tiene actualmente indexadas un total de 764 publicaciones y Scopus 1568, una cantidad no demasiado grande que se reduce especialmente si consideramos solamente las tipologías citables. Sin embargo, no es posible con estas fuentes trazar el crecimiento diario ya que no ofrecen el día exacto de publicación o al menos de inclusión en la base de datos. Sin embargo si se puede limitar la búsqueda a los registros ingresados durante las últimas semanas. Si utilizamos el filtro semanal los dos productos nos ofrecen resultados coincidentes mostrando que el 50% de las publicaciones se han incorporado en las dos últimas semanas, es decir los días comprendidos entre el 30 de Marzo y el 12 de Abril. En el caso de WoS el 32% son de la última semana (6-12 de Abril), una cifra que se reduce al 20% en Scopus. En cualquier caso, si bien estas cifras son groseras y limitadas por la fecha de adición de cada base de datos, ya nos apuntan a un crecimiento importante y muy concentrado en los últimos días en relación al tema del COVID-19. Para aproximarnos con mayor concreción a dicho crecimiento hemos utilizado también *Dimensions* una base de datos mixta con fuentes de información de distinta naturaleza.

**Tabla 1.** Producción científica sobre el COVID-19 en las bases de datos *Scopus* y *Web of Science Core Collection* a fecha de 12/04/2020

| | | Web of Science Core Collection | Scopus |
|---|---|---|---|
| **Total de publicaciones** | **Semana** | **718** | **1568** |
| Añadidas última semana | 6 Abril-12 Abril | 231 – 32% | 310 – 20% |
| Añadidas últimas 2 semanas | 30 de Marzo – 12 Abril | 384 – 53% | 757 – 48% |
| Añadidas últimas 4 semanas | 16 de Marzo – 12 de Abril | 575 – 80% | 1042 – 66% |

Empleando el dataset de *Dimensions* el panorama cambia sustancialmente. Existen un total de 9435 publicaciones científicas incluyéndose tanto publicaciones en revistas como en repositorios (por ejemplo PMC, Arxiv,…) y mezclando, por tanto, contribuciones que han pasado el proceso de revisión con otras que no lo han hecho como el caso de los preprints. Esta base de datos hace las funciones más de agregador de fuentes que de base de datos tradicional. La fusión de múltiples fuentes y la inclusión del campo sobre de la fecha de indexación en *Dimensions* permiten trazar un retrato más preciso del crecimiento diario de la producción científica sobre el COVID-19. En el momento en que se declaró la Emergencia Sanitaria Internacional por la *World Health Organization*, el 30 de Enero, *Dimensions* ya contaba con 340 publicaciones indexadas. En un mes esa cifra se dispara y alcanza un corpus significativo de 1466 publicaciones (24 de February). Desde ese momento el número de publicaciones se duplica con un promedio de 15 días, como ocurrió entre el 14 de marzo y el 30 de marzo cuando se pasan de las 3041 publicaciones a las 6022. Nos encontramos con un crecimiento cuyo mejor ajuste es el modelo de crecimiento exponencial con un valor de $R^2$ de 0,92. En cuanto a la distribución del total de publicaciones hay que señalar que 6549 corresponden a publicaciones en revistas científicas estando 5715 de las mismas indexadas en los productos del NIH *Pubmed* y *Pubmed Central* (PCM). Del conjunto un total de 2677 (31%) corresponden a repositorios en acceso abierto de pre-prints.

**Gráfico 1. Evolución del número de publicaciones acumuladas sobre el COVID-19 en Dimensions clasificadas según su fuente (repositorios y revistas)**

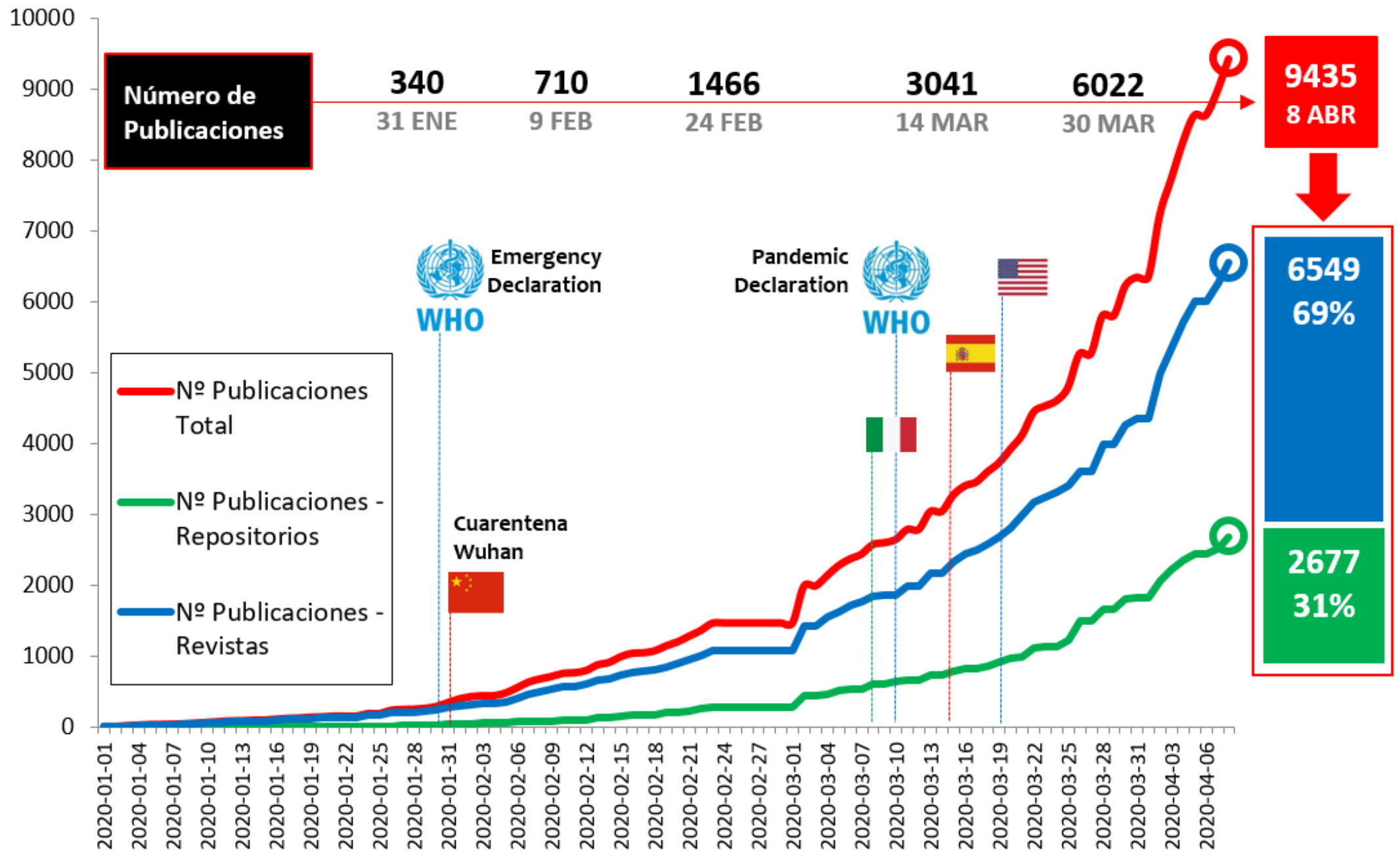

Durante los últimos seis días, comprendidos entre el 2-6 de Abril, se han añadido un promedio de 514 trabajos diarios a *Dimensions*. Para complementar estos datos una de las bases de datos fundamentales es *Pubmed*. Para analizar el número de publicaciones en dicha base de datos hemos tomado como referencia todos aquellos registros del dataset de *Dimensions* que contaban con el identificador Pubmedid. Una vez aislados dichos registros obtenemos un total de 4291 y si las contabilizamos desde el 20 de Enero se reducen a 3844 (Tabla 2). Los datos de *Dimensions* los hemos comparado con los ofrecidos oficialmente por *Litcovid*, una fuente oficial del NIH que extrae y selecciona los artículos más relevantes de *Pubmed*. En *Litcovid* la cifra total de publicaciones es de 3922, por tanto ambas bases de datos recogen un número similar de publicaciones. Aunque en las primeras semanas hay diferencias sustanciales en el número de trabajos indexados a favor *Dimensions* éstas se reducen las últimas semanas, especialmente las comprendidas entre el 23 Marzo-29 Marzo y el 30 Marzo - 5 Abril cuando la diferencia es del 7% y 4% respectivamente. La cobertura similar de dos fuentes diferentes nos permite afirmar que actualmente la producción científica sobre COVID-19 en *Pubmed* se sitúa en torno a las 1000 publicaciones semanales. El crecimiento de la producción reproduce el esquema exponencial ya señalado con los datos del Gráfico 1, en ambos casos el gran salto se produce en la semana del 30 de Marzo al 5 Abril cuando *Dimensions* y *LitCovid* incrementan su número de registros en un 91% y un 97% respectivamente en relación a la semana anterior.

**Tabla 2. Evolución de las publicaciones sobre COVID-19 en la base de datos Pubmed a través de los registros recogidos en Dimensions y en LitCovid**

| | **Número de publicaciones** | | | **Incrementos semanales** | |
|---|---|---|---|---|---|
| **Semana** | **Nº Pubs. Pubmed LitCovid** | **Nº Pubs. Pubmed Dimensions** | **Diferencia a favor de Dimensions** | **Crecimiento porcentual semanal LitCovid** | **Crecimiento porcentual semanal Dimensions** |
| 20 Enero - 26 Enero | 13 | 68 | 81% | | |
| 27 Enero - 02 Febrero | 45 | 103 | 56% | + 246% | + 51% |
| 03 Febrero - 09 Febrero | 102 | 172 | 41% | + 127% | + 67% |
| 10 Febrero - 16 Febrero | 116 | 194 | 40% | + 14% | + 13% |
| 17 Febrero - 23 Febrero | 139 | 235 | 41% | + 20% | + 21% |
| 24 Febrero - 01 Marzo | 173 | 300 | 42% | + 24% | + 28% |
| 02 Marzo - 08 Marzo | 263 | 284 | 7% | + 52% | - 5% |
| 09 Marzo - 15 Marzo | 266 | 358 | 26% | + 1% | + 26% |
| 16 Marzo - 22 Marzo | 406 | 562 | 28% | + 53% | + 57% |
| 23 Marzo - 29 Marzo | 499 | 539 | 7% | + 23% | - 4% |
| 30 Marzo - 5 Abril | 985 | 1029 | 4% | + 97% | + 91% |
| 6 Abril - 12 Abril | 915 | *317 | -- | - 1 % | -- |
| **Totales** | **3922** | **3844** | **--** | -- | -- |
| *Datos disponibles hasta el 8 de Abril | | | | | |

## 3.2. Datos de repositorios y acceso abierto

En el Grafico 2 presentamos los datos el número de publicaciones indexadas en ocho repositorios diferentes. En este análisis hemos incluido los registros de PMC que tiene otras características a los repositorios de pre-prints ya que indexa contenidos en acceso abierto de revistas científicas. Si consideramos en la contabilización a PMC el total de publicaciones en repositorios es de 4075. La lista la lidera la mencionada PCM que acumula 1398 publicaciones (34% del total) con una actividad significativa desde los primeros días de la pandemia. El siguiente repositorio con mayor cobertura es *medRxiv* con 998 (24%) publicaciones que empieza a publicar contenido de forma relevante a partir de marzo. La tercera fuente que contribuye a la producción sobre el COVID-19 es el SSRN con 608 (15%) pre-prints. Todas ellas presentan crecimientos que se acercan al modelo exponencial. Junto a estas fuentes encontramos otros cinco repositorios (*Arxiv*, *bioRxiv*, *Research Square*, *ChemoRxiv* y *JMIR Preprints*) que contribuyen con 1080 publicaciones (27%), todas ellas empiezan a mediados de marzo a publicar contenido de forma relevante y creciente

**Gráfico 2. Evolución del número de publicaciones acumuladas sobre el COVID-19 en ocho repositorios diferentes**

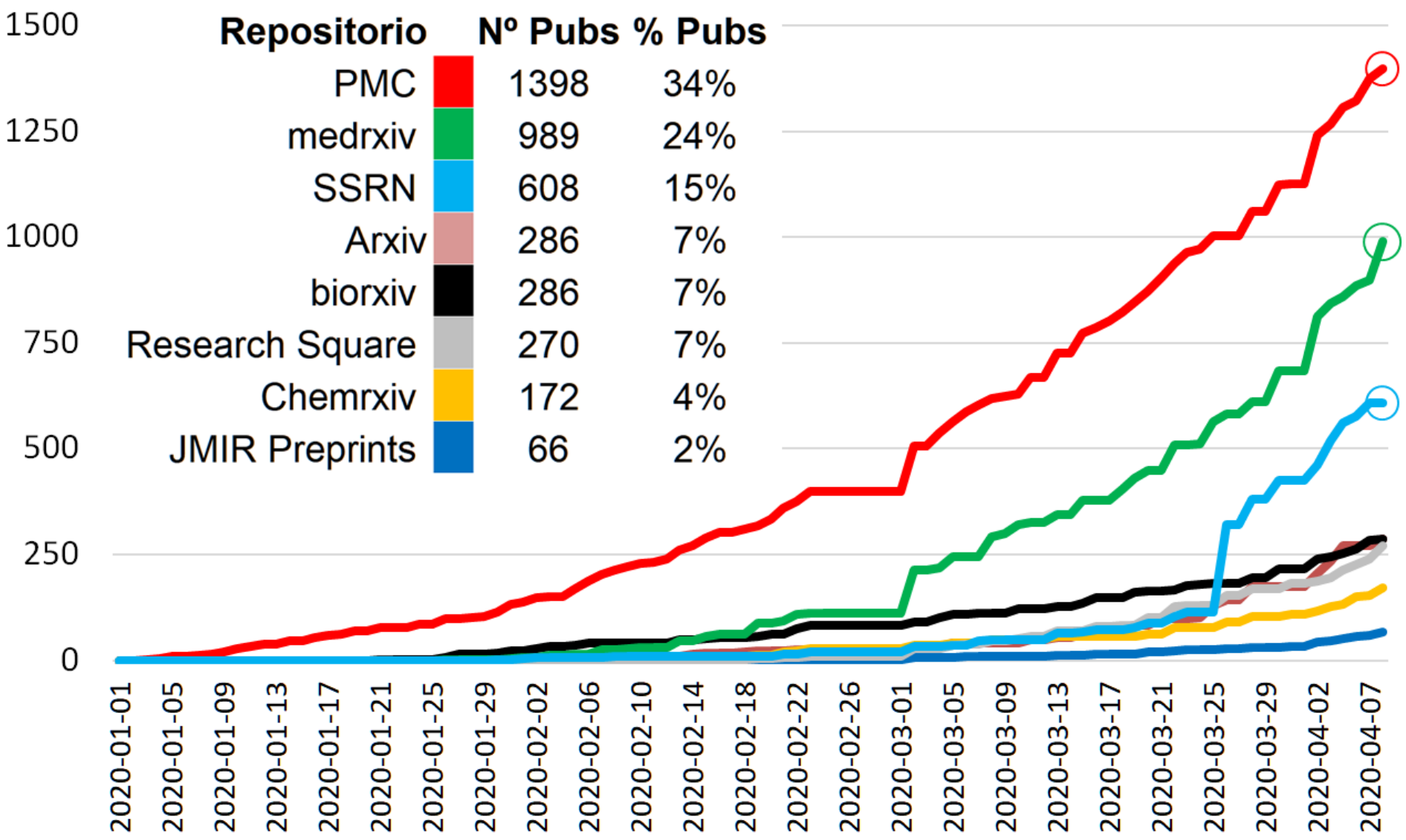

Finalmente consideramos relevante ofrecer las cifras del acceso abierto en el frente de investigación del COVID-19 ya que son espectaculares y singulares. *Dimensions* nos indica que un total de 6357 publicaciones están en acceso abierto, es decir el 67%. Si solo consideramos las publicaciones en revistas científicas encontramos un total de 3832 que porcentualmente es el 60%. La cifra anterior se eleva al 67% si solo consideramos las revistas de Pubmed (2843 de un total 4291). En *Web of Science* el porcentaje de artículos en Acceso Abierto se incrementa al 97% ya que casi todas las grandes revistas científicas ofrecen abiertos su contenidos sobre el tema del COVID-19, tal y como indicamos en la introducción. En Scopus las cifras son también elevadas y el porcentaje de publicaciones en acceso abierto es del 84%. Esto nos indica que aproximadamente tres de cada cuatro publicaciones que aparecen en revistas sobre el Covid 19 están en acceso abierto a las que habría que sumar todos los preprints de los repositorios.

## 4. Consideraciones finales

En relación a los resultados obtenidos se ha determinado el ritmo de crecimiento que a nivel global es de 500 publicaciones diarias en *Dimensions*. En el caso de *Pubmed* el crecimiento semanal se sitúa en torno a las 1000 publicaciones. Por tanto nos encontramos ante un fenómeno clásico bibliométrico de crecimiento exponencial ($R^2$ =0,92). Desde el punto de vista de la bases de datos, los modelos más tradicionales de bases de datos se ven perjudicados ya que a veces tardan en incorporar los registros y pueden que no sean los agiles e inmediatos que requieren un tema como el COVID-19. *Dimensions* ha resultado ser un recurso más exhaustivo al integrar diversas fuentes y por tanto lo convierten en una fuente más valiosa desde el punto de vista informativo. Asimismo como no podía de otra manera *Pubmed* se mantiene como la gran base de datos de referencia con una respuesta en cobertura y acceso considerable.

En relación al acceso abierto en sus dos vertientes está jugando un papel fundamental. Por un lado los editores científicos se han volcado con la puesta disposición de los trabajos sobre el COVID-19, alcanzándose porcentajes de acceso abierto inéditos en revistas. Los repositorios también están ocupando un lugar central si bien no sustituyen a las revistas científicas ya que la  publicaciones de la modalidad ruta verde suponen solo el 30% de las publicaciones. Este estudio ha contribuido, por tanto, a tener una imagen amplia sobre el volumen de la producción científica del COVID-19 y sus fuentes más relevantes. Hemos de terminar reseñando que el reto actual no solo atañe al ámbito médico sino que también exige una respuesta de los profesionales de la información ante una explosión de información sin precedentes que nos sitúa en el centro de la pandemia informativa.

## 5. Bibliografía

Alba-Ruiz, Rubén. (2020). "COVID-19, CORONAVIRUS PANDEMIC: aproximación bibliométrica y revisión de los resultados". Zenodo, 31 Marzo
http://doi.org/10.5281/zenodo.3734062

Chahrour, M.; Assi, S.; Bejjani, M.; Nasrallah, A. A.; Salhab, H.; Fares, M. Y., & Khachfe, H. H. (2020). "A Bibliometric Analysis of COVID-19 Research Activity: A Call for Increased Output". Cureus, v.12, n. 3, e7357
http://doi.org/10.7759/cureus.7357

Chen, Q.; Allot, A & Lu, Z.  (2020). "Keep up with the latest coronavirus research". Nature. v. 579, n. 7798, pp. 193.
http://dx.doi.org/10.1038/d41586-020-00694-1

Hossain, Md Mahbub (2020). "Current Status of Global Research on Novel Coronavirus Disease (COVID-19): A Bibliometric Analysis and Knowledge Mapping". SSRN, 2 Abril
http://dx.doi.org/10.2139/ssrn.3547824